\documentclass[aps,prl,twocolumn,superscriptaddress,preprintnumbers,amsmath,amssymb,longbibliography,10pt]{revtex4-2}
\usepackage{dcolumn}
\usepackage{graphicx}
\usepackage{amsmath,amsfonts,amssymb,amsthm,amstext,amsbsy,amsopn,amscd,amsxtra}
\usepackage{dsfont}
\usepackage{physics} 
\usepackage{upgreek} 
\usepackage[colorlinks=true,allcolors=blue]{hyperref}
\usepackage{xcolor}
\usepackage{url}
\usepackage{siunitx}
\usepackage{multirow}
\usepackage{hhline}
\usepackage{bm}

\everymath{\displaystyle}

\newcommand{\dL}{\text{L}}
\newcommand{\Fr}{\text{F}_\text{rat}}
\newcommand{\dc}{\text{c}}
\newcommand{\dQOC}{\text{QOC}}
\newcommand{\deff}{\text{eff}}

\begin{document}

\title{A Hamiltonian ratchet for matter wave transport}

\author{N. Dupont}
\affiliation{
Laboratoire Collisions Agr\'egats R\'eactivit\'e, UMR 5589, FERMI, UT3, Universit\'e de Toulouse, CNRS,\\
118 Route de Narbonne, 31062 Toulouse CEDEX 09, France
}
\affiliation{Center for Nonlinear Phenomena and Complex Systems,
Universit\'e Libre de Bruxelles, CP 231, Campus Plaine, B-1050 Brussels, Belgium}
\author{ L. Gabardos}
\affiliation{
Laboratoire Collisions Agr\'egats R\'eactivit\'e, UMR 5589, FERMI, UT3, Universit\'e de Toulouse, CNRS,\\
118 Route de Narbonne, 31062 Toulouse CEDEX 09, France
}
\author{ F. Arrouas}
\affiliation{
Laboratoire Collisions Agr\'egats R\'eactivit\'e, UMR 5589, FERMI, UT3, Universit\'e de Toulouse, CNRS,\\
118 Route de Narbonne, 31062 Toulouse CEDEX 09, France
}
\author{ N. Ombredane}
\affiliation{
Laboratoire Collisions Agr\'egats R\'eactivit\'e, UMR 5589, FERMI, UT3, Universit\'e de Toulouse, CNRS,\\
118 Route de Narbonne, 31062 Toulouse CEDEX 09, France
}
\author{J. Billy}
\affiliation{
Laboratoire Collisions Agr\'egats R\'eactivit\'e, UMR 5589, FERMI, UT3, Universit\'e de Toulouse, CNRS,\\
118 Route de Narbonne, 31062 Toulouse CEDEX 09, France
}
\author{B. Peaudecerf}
\affiliation{
Laboratoire Collisions Agr\'egats R\'eactivit\'e, UMR 5589, FERMI, UT3, Universit\'e de Toulouse, CNRS,\\
118 Route de Narbonne, 31062 Toulouse CEDEX 09, France
}
\author{D. Gu\'ery-Odelin}
\affiliation{
Laboratoire Collisions Agr\'egats R\'eactivit\'e, UMR 5589, FERMI, UT3, Universit\'e de Toulouse, CNRS,\\
118 Route de Narbonne, 31062 Toulouse CEDEX 09, France
}
\email{dgo@irsamc.ups-tlse.fr}

\date{\today}

\begin{abstract}

We report on the design of a Hamiltonian ratchet exploiting periodically at rest integrable trajectories in the phase space of a modulated periodic potential, leading to the linear non-diffusive transport of particles. Using Bose-Einstein condensates in a modulated one-dimensional optical lattice, we make the first observations of this spatial ratchet, which provides way to coherently transport matter waves with possible applications in quantum technologies. 
In the semiclassical regime, the quantum transport strongly depends on the effective Planck constant due to Floquet state mixing. We also demonstrate the interest of quantum optimal control for efficient initial state preparation into the transporting Floquet states to enhance the transport periodicity.
\end{abstract}

\maketitle

\begin{figure}[t]
\begin{center}
\includegraphics[scale=1]{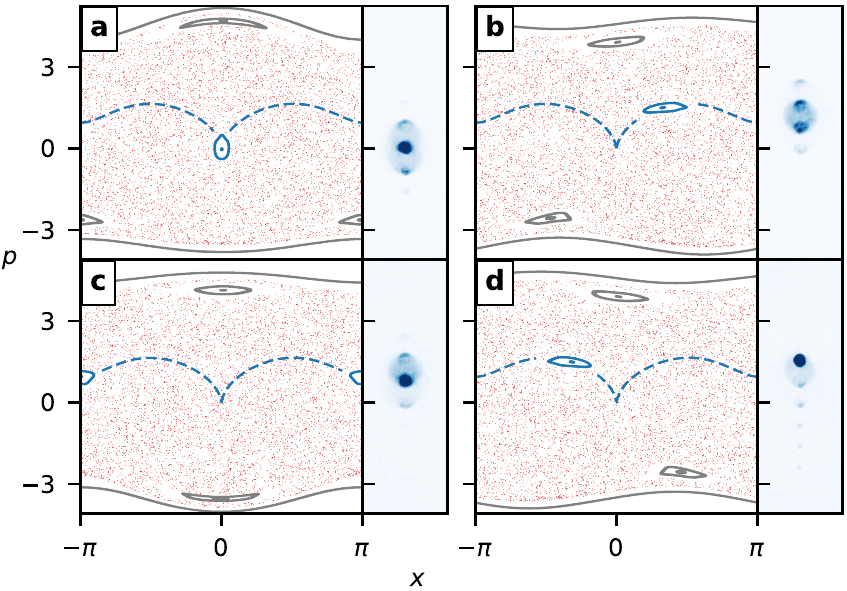}
\caption{
\textbf{Stroboscopic phase portraits and experimental images} for $(\gamma, \varepsilon, \varphi_0) = (1.2, 0.3, 1.7)$ at sub-period observation times $t=(n+r)\times 2 \pi$ (with $n \in \mathbb{N}$ and $r=0$, $0.25$, $0.5$ and $0.75$ for \textbf{(a)} to \textbf{(d)} resp.). Left: the ratcheting island and the trajectory starting in $(x_0, p_0, t_0) = (0,0,0)$ are in blue, the other regular structures are in gray and the chaotic sea is in red. The area of the ratcheting island is $\mathcal{A} = 0.21$. Right: corresponding time-of-flight absorption images, starting from the ground state of the lattice during the first period $n=0$ for $1/\hbar_\deff \approx 1.27$.
}
\label{fig1}
\end{center}
\end{figure}

The ratchet effect is the well-known yet intriguing phenomenon which sees the emergence of a directed current for particles initially at rest in a space and time periodic modulated potential, while no net average force is exerted on the system.
Its origin is well understood and relies minimally on the breaking of space and time reversal symmetries~\cite{flach_2000, denisov_2002, denisov_2014}. 
Two main families of ratchets can be distinguished:
on one hand, \textit{Brownian ratchets} are systems experiencing stochastic forces, where the potential rectifies the isotropy of Brownian motion \cite{hanggi_2005} into a net directed transport~\cite{rousselet_1994, mr_1999, astumian_2002, schiavoni_2003, sanchez_2004, sjolund_2006, hanggi_2009}. 
Such ratchets
are thought to 
take part in the 
operation of 
molecular motors \cite{julicher_1997,chowdhury_2005}, as for instance in the case of kinesin \cite{nishiyama_2002}. They are usually studied in the overdamped regime to model the strong dissipation of biological media \cite{reimann_2002, hanggi_2009}. 
On the other hand, \textit{deterministic ratchets}, which can be either dissipative~\cite{jung_1996, flach_2000, carlo_2005, wang_2007, denisov_2009} or Hamiltonian (see below), are systems for which the  classical dynamics is well captured by their phase space flow. Such Hamiltonian systems, under moderate temporal driving, exhibit a mixed dynamics with phase portraits displaying islands of regular trajectories embedded in a chaotic sea of 
non-integrable ones (see \textit{e.g.} Fig.~\ref{fig1}). 

Studies on Hamiltonian ratchets have mainly focused thus far on delocalized transport configurations, where the directed transport originates either in a momentum-asymmetic chaotic sea (in the classical case from trajectories ergodically spanning the chaotic sea~\cite{flach_2000, schanz_2001, schiavoni_2003}, and in the quantum case through state coupling with eigenstates delocalized over it~\cite{denisov_2007, schanz_2001, salger_2009, zhan_2011}), or by resonantly accelerating particles 
in the case of quantum-resonance ratchets~\cite{lundh_2005,ni_2017}.
In contrast, Hamiltonian ratchets relying on regular islands of quasi-periodic trajectories offer a mean to incrementally transport localized particles on a periodic substrate in a ballistic way~\cite{denisov_2007}. Note that this transport appears classically and is distinct from topological pumping effects (\cite{Citro_2023} and references therein).
Such regular Hamiltonian ratchets have been experimentally studied mainly with phase-shifted kicked-rotors, implemented in cold atom systems, 
in the case of step-wise transport along the momentum direction (an \emph{accelerator ratchet})~\cite{gong_2004, sadgrove_2013, white_2013, hainaut_2018}, and with only up to 20\% of an initial atom packet loaded in the transporting island.
Meanwhile, regular Hamiltonian ratchets along the position coordinate~\cite{dittrich_2000} remain unexplored experimentally so far. Beyond the use of regular islands with a ballistic motion at all times, as found \textit{e.g.} in the kicked rotor, of particular interest is the design of a dynamical system in which the transporting island periodically coincides with the ground state of the potential, in which a collection of particles initially at rest can therefore be directly loaded and transported.

In this Letter, we solve this non-trivial problem, the solution of which we refer to as a \emph{spatial halting ratchet} (SHR). We show how such a 
solution can be engineered with a simple gating ratchet \cite{borromeo_2005, gommers_2008}, a 
one-dimensional 
space-symmetric
potential modulated in amplitude and phase. We obtain parameters leading to classical orbits of initial zero velocity having a ratcheting motion of one spatial period per modulation period. For these parameters, we study quantum transport as a function of the effective reduced Planck constant $\hbar_\deff$,
a free quantum parameter that sets the minimal phase-space area of quantum states without altering the classical dynamics.
As a modulation parameter (\textit{e.g.} $\hbar_\deff$) is varied, 
avoided crossings in the Floquet spectrum lead to state mixing. 
Although this phenomenon is sought after in the case of diffusive Hamiltonian ratchets~\cite{salger_2009, grossert_2016} where it is the source of transport, and is also key in chaos-assisted tunneling~\cite{tomsovic_1994, arnal_2020}, it has a deleterious effect in
the 
regular transport case, as 
a wave packet initially prepared over the classical ratcheting island  
may dynamically tunnel~\cite{book_keshavamurthy_2011} out of it. This may be avoided by a precise choice of parameter values, or controlled through specific state preparation.
From our theoretical analysis, we implement and observe experimentally the SHR with matter waves, using Bose-Einstein condensates (BECs) in a modulated one-dimensional optical lattice.
We first perform these experiments by loading the ratcheting Floquet state from the ground state of the lattice. For values of $\hbar_\deff$ for which the ratcheting island is substantially coupled with the chaotic sea, we then show how one can account for state mixing by employing quantum optimal control (QOC)~\cite{boscain_2021, koch_2022} to optimize the loading of the proper Floquet state.

\paragraph*{Classical dynamics.} We consider the case of an inertial particle in a gating potential~\cite{borromeo_2005, gommers_2008}. Its dynamics is governed by the dimensionless Hamiltonian
\begin{equation}
H(x,p,t) = \dfrac{p^2}{2} - \gamma \left[1+\varepsilon \cos (t)\right] \cos\left[x - \varphi_0 \sin (t) \right].
\label{eq:gating_hamiltonian_adim}
\end{equation}
\noindent The Hamiltonian of Eq.~\eqref{eq:gating_hamiltonian_adim}, with its 1:1 frequency ratio and phase quadrature between amplitude and phase modulations, breaks the relevant space and time symmetries~\cite{gommers_2008, flach_2000}, leading to a momentum-asymmetric chaotic sea carrying diffusive ratchet transport. In contrast, the dimensionless modulation parameters $(\gamma, \varepsilon, \varphi_0)$ can also be chosen so that a transporting regular region emerges in the center of the chaotic sea, \emph{i.e.} such that a bundle of trajectories starting around $(x_0,p_0)=(0,0)$ at $t_0=0$ goes to the neighborhood of $(x_0+2\pi,p_0)$ at $t=2\pi$.  
We achieve this numerically by minimizing with respect to $(\gamma, \varepsilon, \varphi_0)$ (using a Nelder-Mead algorithm) the total variation of mechanical energy over one modulation period for a set of trajectories that start near $(x_0, p_0)$ and change site. This yields several solutions~\cite{DupontPhD}. 
In the following we use  $(\gamma, \varepsilon, \varphi_0) = (1.2, 0.3, 1.7)$, a set of parameters that generates a SHR with the ratcheting island seen in the 
stroboscopic phase portraits 
of Fig.~\ref{fig1}.

\paragraph*{Quantum ratchet in a regular island.} The natural basis to stroboscopically study quantum dynamics in a time-periodic potential is the set of Floquet states, the eigenstates of the evolution operator over one period of modulation. The quantum study leaves as a free parameter the effective reduced Planck constant $\hbar_\deff = -i[\hat{x},\hat{p}] $ that dictates the minimal phase-space area $\Delta x \Delta p$ of quantum states in the system. 
As we are interested in the transport of a quantum particle on the ratcheting island, we place our study 
at the onset of the semiclassical regime, that is for $\hbar_\deff \sim \mathcal{A}$, with $\mathcal{A}$ the area of the studied regular structures in phase space (Fig.~\ref{fig1}).
In the semiclassical regime, Floquet states are generally either localized on 
regular islands or spread over the chaotic sea~\cite{berry_1984, bohigas_1993}, with only one state per island for $\hbar_\deff \sim \mathcal{A}$. 
The quantum analogue of the periodic classical trajectories at the center of the stroboscopic phase portraits of Fig.~\ref{fig1}(a) is therefore the Floquet state $\ket{\Fr}$ that can be associated with the ratcheting island. This state is identified from its overlap with the ground state $\ket{\phi_0}$ of the static lattice potential (that is $\varepsilon=0$ and $\varphi_0=0$ in Eq.~\eqref{eq:gating_hamiltonian_adim} ; a state readily accessible in the experiment).
Furthermore, we define the expected transport of a state $\ket{\psi(t_0)}$ between the times $t_0$ and $t_1$ 
as 
\begin{equation}
\Delta x_{(t_0,t_1)}(\psi) = \int_{t_0}^{t_1} \left\langle \hat{p} \right\rangle_{\psi(t)} \, dt.
\label{eq:expected_transport}
\end{equation} 
\noindent  
The transport over one period $\Delta x_{(0,2\pi)}(\mathrm{F})$ for a Floquet state $\ket{\mathrm{F}}$ is related to its time-averaged group velocity $\overline{v}_\text{g}$ in the Floquet spectrum,  $\Delta x_{(0,2\pi)}(\mathrm{F})=2\pi \overline{v}_\text{g}$ (see Supplemental Material).
In the semiclassical regime, one expects regular Floquet states to behave as their associated region of regular classical trajectories, and, in particular for the ratcheting Floquet state, $\Delta x_{(0,2\pi)}(\Fr) \approx 2\pi$.
Therefore in that regime, the very existence of a transporting regular island generally guarantees for all quasi-momenta (see below and Supplemental Material) the existence of a localized transporting Floquet state, a key feature differing from previous ballistic ratchets~\cite{salger_2009, zhan_2011}.

\begin{figure}[t]
\begin{center}
\includegraphics[scale=1]{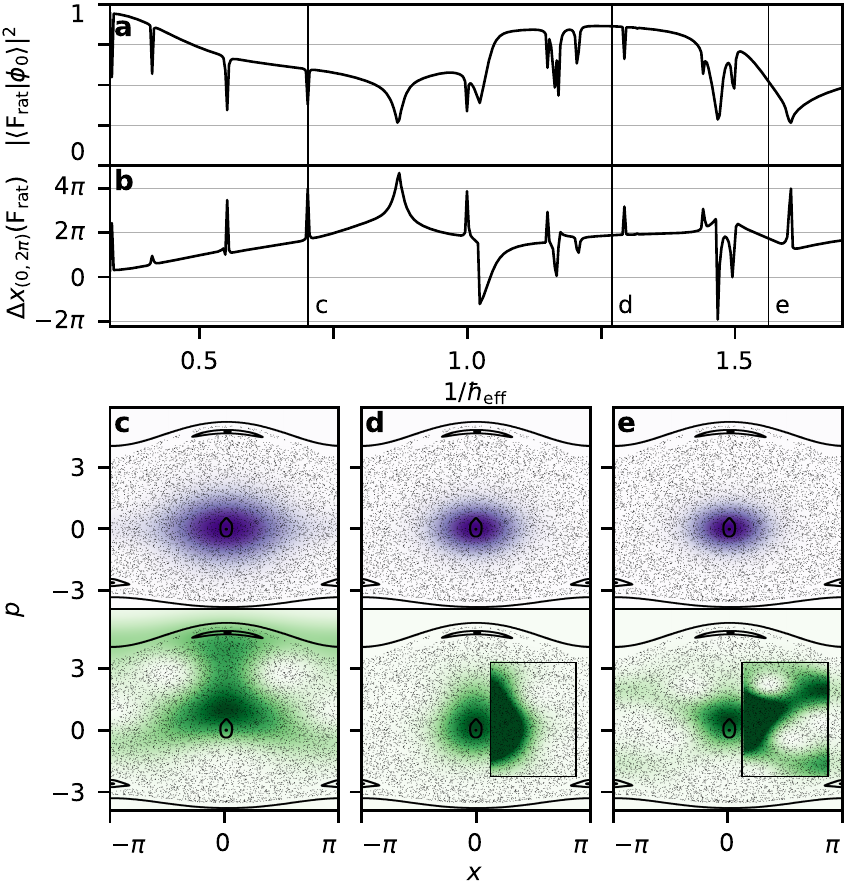}
\caption{
\textbf{Eigenstate and transport dependences on the effective Planck constant.} \textbf{(a)} Overlap between the ground state $\ket{\phi_0}$ of the lattice and the ratcheting Floquet state $\ket{\Fr}$ and \textbf{(b)} transport~\eqref{eq:expected_transport} of $\ket{\Fr}$ over one modulation period as a function of $1/\hbar_\deff$. \textbf{(c-e)} Stroboscopic phase portraits in the unit cell of system~\eqref{eq:gating_hamiltonian_adim} and Husimi representations of $\ket{\phi_0}$ (top, purple) and $\ket{\Fr}$ (bottom, green) for the values of $1/\hbar_\deff=0.70,1.27,1.56$ respectively, identified by vertical lines on the panels (a-b). 
The color range for each Husimi function extends from zero to its maximum value, with a truncation to a quarter of this value in the outlined rectangular regions of panels (d,e) in order to reveal details (note that the Floquet states share the $x\rightarrow-x$ symmetry of the phase portrait).
}
\label{fig2}
\end{center}
\end{figure}

We illustrate these notions in Fig.~\ref{fig2} where numerical results for the transport of non-interacting wave functions in the ratcheting island 
as a function of $1/\hbar_\deff$ are shown.
Figure~\ref{fig2}(a) shows the overlap between $\ket{\Fr}$ and the ground state $\ket{\phi_0}$. 
This metric informs on the phase-space centering of $\ket{\Fr}$, as well as on its expected loading when running the experiment with $\ket{\phi_0}$ as the initial state. Figure~\ref{fig2}(b) shows 
the expected periodic transport of $\ket{\Fr}$. 
As $1/\hbar_\deff$ varies, both $|\langle \phi_0 | \Fr \rangle|^2$ and $\Delta x_{(0,2\pi)}(\Fr)$ (Fig.~\ref{fig2}(a) and (b) resp.) display sharp non-monotonic fluctuations ascribed to Floquet state mixing: the variation of quasi-energy levels in the Floquet spectrum gives rise to avoided crossings leading to sharp changes of the Floquet states near the crossings (see Supplemental Material).

Figures~\ref{fig2}(c-e) shows the Husimi quasi-distributions of $\ket{\phi_0}$ and $\ket{\Fr}$ for different values of $1/\hbar_\deff$. At given coordinates $(x,p)$, this phase space representation of a quantum state corresponds to the evaluation of its overlap with a Gaussian state centered in $(x,p)$~\cite{bahr_2007, dupont_2023}. 
With $1/\hbar_\deff = 1.27$, Fig.~\ref{fig2}(d) is an example of a semiclassical, island-shaped $\ket{\Fr}$, with
a ground state overlap
$|\langle \phi_0| \Fr \rangle|^2 = 0.86$ 
and a periodic transport $\Delta x_{(0,2\pi)}(\Fr) = 0.93\times2\pi$, meaning that this state exhibits a stationary flux of particles, periodically at rest at the bottom of the lattice wells. On the other hand, Fig.~\ref{fig2}(c) and (e) correspond to respectively smaller and larger values of $1/\hbar_\deff$ for which the system initialized in $|\phi_0\rangle$ evolves out of it (towards a mode of high momentum for Fig.~\ref{fig2}(c) and over the chaotic sea for Fig.~\ref{fig2}(e)). 
Figure~\ref{fig2} shows overall that, for sufficiently large values of $1/\hbar_\deff$ and while paying attention to Floquet state mixing, a SHR with a semiclassical periodic transport of quantum states can be achieved, with $\Delta x_{(0,2\pi)}(\Fr)$ fluctuating around $2\pi$ for $1/\hbar_\deff > 0.75$.

\paragraph*{Ratchet transport from the ground state.} We experimentally observe a SHR with matter waves using BECs of $5 \cdot 10^5$ $^{87}$Rb atoms initially obtained in a hybrid trap setup~\cite{fortun_2016}. 
The atoms are adiabatically loaded at time $T_0=0$ in the ground state $\ket{\phi_0}$ of the optical lattice potential
\begin{equation}
V(X,T) = -A(T)\dfrac{s}{2}E_\dL \cos(\frac{2\pi X}{d} + \varphi(T)),
\label{eq:lattice_potential}
\end{equation}
\noindent with $A(T_0) = 1$ and $\varphi(T_0) = 0$ (we denote with capital $X$, $P$ and $T$ dimensional quantities). The optical lattice is produced by the superposition of two counterpropagating far-detuned laser beams of wavelength $\lambda = 1064$ nm. Before each experiment, we independently calibrate~\cite{cabrera_2018} the depth $s$ of the lattice in units of the lattice energy scale $E_\dL = h^2 / 2md^2$ (with $d=\lambda/2$ the lattice spacing, $m$ the atomic mass and $h$ Planck's constant). The driving amplitude of an acousto-optic modulator (AOM) placed before the splitting of the lattice beams controls $A(T)$, while the relative driving phase of two AOMs following the beams splitting controls $\varphi(T)$. 
The optical lattice potential~\eqref{eq:lattice_potential}, with the correlated modulation functions $A(T) = (1+ \varepsilon \cos(\omega T))$ and $\varphi(T) = - \varphi_0 \sin(\omega T)$ where $\omega$ is the modulation angular frequency,
yields 
the dimensionless 
gating Hamiltonian~\eqref{eq:gating_hamiltonian_adim}  for $\gamma = s (E_\dL/\hbar \omega)^2$ and an effective reduced Planck constant $\hbar_\deff  = 2E_\dL/\hbar \omega$ \footnote{Having $\gamma = s (E_\dL/\hbar \omega)^2$ and $\hbar_\deff = 4\pi^2 \hbar / m\omega d^2$, 
$\hbar_\deff$  is varied without altering the classical dynamics by adjusting the lattice depth $s = 4\gamma / \hbar_\deff^2$.}. 

The $1/\hbar_\deff$ range of Fig.~\ref{fig2} corresponds in practice to a lattice depth range $s\in[0.5, 14]$.
A weak harmonic trapping 
with angular frequencies $(\Omega_X,\Omega_Y,\Omega_Z) = 2\pi \times (10.4, 66,68)$ Hz remains present during experiments, but its impact is negligible  
over the short experimental times of up to $\sim500$~$\upmu$s in this work. In the subspace of null quasi-momentum, the BEC state along the $x$-axis is thus described by a superposition of plane waves
\begin{equation}
\ket{\psi(T)} = \sum_{\ell\in\mathbb{Z}} c_\ell(T) \ket{\chi_\ell},
\label{eq:BEC_state}
\end{equation}
\noindent with the coefficients $c_\ell(T) \in \mathbb{C}$, $\textstyle \sum_\ell |c_\ell(T)|^2 = 1$ and $\langle X | \chi_\ell \rangle = e^{i \ell k_\dL X}/ \sqrt{d}$, where $k_\dL=2\pi/d$ is the lattice wavevector.
Finally, we access at time $T$ the BEC momentum distribution by absorption imaging following a 35\,ms time-of-flight.
We obtain the typical diffraction patterns of Fig.~\ref{fig1},
from which we extract $|c_\ell(T)|^2$. 
The experimental transport~\eqref{eq:expected_transport} is then computed by sampling the average momentum $\textstyle \langle\hat{P}\rangle_{\psi(T)}/\hbar k_\dL = \sum_\ell \ell |c_\ell(T)|^2$ in the course of the ratchet modulation. In this system, a transport of 1 site per modulation period corresponds to a velocity $v = \hbar k_\dL/m\hbar_\deff \approx 8.63/\hbar_\deff$ mm/s.

\begin{figure}[t]
\begin{center}
\includegraphics[scale=1]{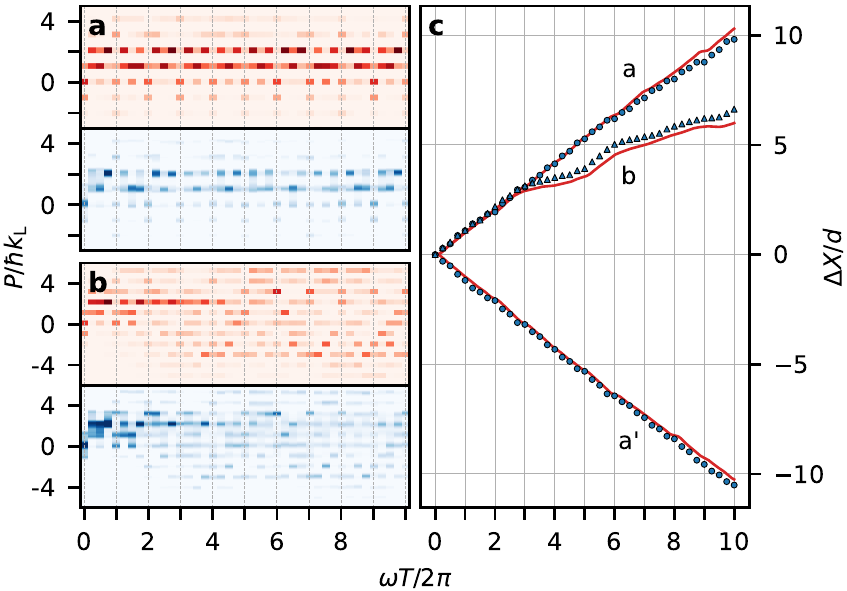}
\caption{
\textbf{Transport of the ground state.} \textbf{(a)} Top: Numerical simulation of the momentum distribution during the modulation as a function of time for $1/\hbar_\deff \approx 1.27$ (corresponding to Fig.~\ref{fig2}(d)). Bottom: Corresponding experimental integrated absorption images. \textbf{(b)} Same as (a) for $1/\hbar_\deff \approx 1.56$ (corresponding to Fig.~\ref{fig2}(e)). \textbf{(c)} Expected numerical (solid red line) and experimental (blue markers) transport (see text) for data (a) and (b) as a function of time, and transport reversability (a') for $\varphi_0 \rightarrow -\varphi_0$ and $1/\hbar_\deff \approx 1.30$ (see text).
}
\label{fig3}
\end{center}
\end{figure}

We first perform ratchet transport experiments with $\ket{\phi_0}$ as the initial state. We acquire 4 images per modulation period over 10 periods (as shown in Fig.~\ref{fig1} for the first period). Figure~\ref{fig3} shows, for two values of $1/\hbar_\deff$, the experimental evolution of the momentum distribution and the resulting integrated transport compared with numerical simulations of the same quantities from the integration of Schr\"odinger equation. The values of $1/\hbar_\deff$ in Fig.~\ref{fig3}(a) and (b) correspond to those of Fig.~\ref{fig2}(d) and (e) respectively.
For Fig.~\ref{fig3}(a), $\ket{\phi_0}$ is rather well projected onto $\ket{\Fr}$ (Fig.~\ref{fig2}(a,d)). We thus observe an almost periodic evolution of the momentum distribution, mainly carried by plane waves of positive momentum and resulting in a linear semiclassical ratchet transport over 10 lattice sites in 10 modulation periods (Fig.~\ref{fig3}(c)~; a study of the transport over longer timescales is detailed in the Supplemental Material).
In the experiment of Fig.~\ref{fig3}(b) however, as $\ket{\phi_0}$ has limited overlap with $\ket{\Fr}$, which moreover extends over the chaotic sea (Fig.~\ref{fig2}(e)), we observe a non-periodic evolution of the momentum distribution associated with a diffusion over the chaotic sea as seen from the increase of the momentum dispersion. This results in a non-linear evolution of the transport (Fig.~\ref{fig3}(c)), in contrast to its classical counterpart.
Interestingly, this non-classical behavior happens for a 
smaller value of $\hbar_\deff$, 
highlighting the quantum nature of the underlying mechanism of state mixing.

A key feature of the ratchet effect is the ability to reverse the transport direction via adequate symmetries~\cite{hanggi_2009, denisov_2014}. In our gating system, this transport direction is imposed by the sign of the phase quadrature between the amplitude and phase modulations (Eq.~\eqref{eq:gating_hamiltonian_adim}). The change $\varphi(t) = - \varphi_0 \sin(t) \rightarrow + \varphi_0 \sin(t)$ is thus expected to result in a reversed ratchet transport in the lattice. In Fig.~\ref{fig3}(c), the integrated transport for $(\gamma, \varepsilon, \varphi_0) = (1.2, 0.3, -1.7)$ and a value of $\hbar_\deff$ similar to that of panel (a) is shown (with label (a')). We measure as expected a symmetric ratchet transport over $-10$ sites in 10 modulation periods. We get an overall excellent agreement between experiments and simulations.

\begin{figure}[t]
\begin{center}
\includegraphics[scale=1]{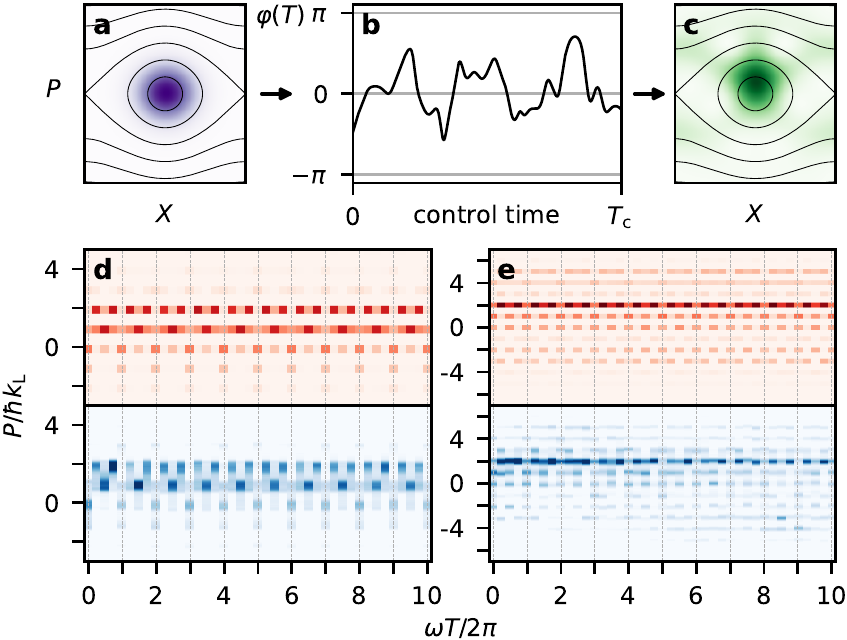}
\caption{
\textbf{Transport of the ratcheting Floquet state prepared by QOC.} \textbf{(a-c)} Example of QOC for ratchet transport at $1/\hbar_\deff \approx 1.56$. \textbf{(a)} Husimi representation of $\ket{\phi_0}$ (purple) in the phase space of the static lattice (solid black lines). The color range for the Husimi function extends from zero to its maximum value. \textbf{(b)} Phase of the lattice along time to drive the system from $\ket{\phi_0}$ to $\ket{\Fr}$. \textbf{(c)} Same as (a) for the prepared state $\ket{\psi_\dQOC}$. \textbf{(d)} Top: Numerical simulation of the momentum distribution during the modulation as a function of time for $1/\hbar_\deff \approx 1.27$ (corresponding to Fig.~\ref{fig2}(d) and Fig.~\ref{fig3}(a)). Bottom: Corresponding experimental integrated absorption images. \textbf{(e)} Same as (d) for $1/\hbar_\deff \approx 1.56$ (corresponding to panels (a-c) as well as Fig.~\ref{fig2}(e) and Fig.~\ref{fig3}(b)).
}
\label{fig4}
\end{center}
\end{figure}

\paragraph*{Optimized loading through quantum optimal control.}  
Even for values of $1/\hbar_\deff$ for which $\Delta x_{(0,2\pi)}(\Fr) \approx 2\pi$,
semiclassical ratchet transport can be limited when working with $|\phi_0\rangle$ as the initial state (see \textit{e.g.} Fig.~\ref{fig2}(a,b,e)).  
To enhance this transport, we use, in a second set of experiments, the phase of the lattice $\varphi$ as a control parameter to optimally prepare $\ket{\Fr}$ before applying the ratchet modulation.
To that end, after determination of $|\Fr\rangle$,
an optimal phase variation $\varphi(0<T<T_\dc)$ in the lattice of fixed depth $s$ is computed using a first-order gradient-ascent algorithm (detailed with its experimental implementation in previous works~\cite{dupont_2021, dupont_2023}), to drive the BEC from $\ket{\phi_0}$ to $\ket{\Fr}$. We set in this work $T_\dc \approx 80\,\mathrm{\mu s}$.
The QOC algorithm converges to a control field that theoretically prepares a state $\ket{\psi_\dQOC}$ with a fidelity of $|\langle \Fr | \psi_\dQOC \rangle|^2 \geq 0.995$~\footnote{This fidelity figure is set as a convergence condition on the iterative QOC algorithm~\cite{dupont_2021}.}, while the fidelities experimentally reached for such targets are typically $\sim 0.95$ ~\cite{dupont_2023}.
We illustrate in Fig.~\ref{fig4}(a-c) the QOC protocol of Floquet state preparation, with an optimized $\varphi(T)$ driving $|\phi_0\rangle$ to $|\Fr\rangle$ for a given value of $1/\hbar_\deff$ (corresponding to Fig.~\ref{fig2}(e) and Fig.~\ref{fig3}(b)).  
Figure~\ref{fig4}(d,e) shows experimental results and numerical simulations for the same parameters as Fig.~\ref{fig3}(a,b) respectively, with a preliminary QOC preparation applied.
While the experiment of Fig.~\ref{fig3}(a) already featured a clear linear quantum transport, Fig.~\ref{fig4}(d) demonstrates how the QOC preparation of the ratcheting Floquet state enhances the periodicity of the momentum evolution. Comparing Fig.~\ref{fig4}(e) with Fig.~\ref{fig3}(b), the gain is even more spectacular. Interestingly, Fig.~\ref{fig4}(e) display a broad momentum dispersion from the beginning of the modulation, which demonstrates the preparation of a ratcheting Floquet state partially extending over the chaotic sea as expected (see Fig.~\ref{fig2}(e)).

\paragraph*{Conclusion}
In this Letter, we have studied a spatial Hamiltonian ratchet effect exploiting regular trajectories in phase space to transport particles periodically at rest. 
We showed how such a SHR can be obtained classically within a gating ratchet. We then considered quantum transport in the near semiclassical regime, for small but realistic values of the effective Planck constant $\hbar_\deff$,
and discussed how quantum transport can be 
strongly affected by Floquet state mixing as $\hbar_\deff$ varies.
We experimentally observed coherent SHR transport of matter waves with BECs of $^{\mathrm{87}}$Rb in a modulated optical-lattice. For values of $\hbar_\deff$ coupling the ratcheting island with the chaotic sea, we witnessed how atoms loaded in the island evolve out of it through dynamical tunneling. 
Finally, we demonstrated how this effect can be mitigated through the use of state control methods such as QOC, to prepare the ratcheting Floquet state and thus enhance the periodicity of the dynamics. 

Our modeling relies on an infinite lattice description, and is in good agreement with experimental data. Finite-size effects therefore have a limited impact on our experiments, which is due both to the extension of the BEC ($\approx100$ lattice sites) and to the fact that in the cases studied here the ratcheting Floquet state has a uniform group velocity in the vicinity of zero quasi-momentum (see Supplemental Material). 

The regular ratchet effect we demonstrated constitutes a novel way to coherently transport matter waves in a periodic potential, alongside conveyor belt approaches~\cite{schrader_2001, hauck_2021, klostermann_2022}. Higher SHR currents could be obtained by decreasing $\hbar_\deff$ in a deeper lattice. Our results lend themselves to further investigations, such as the extension to higher dimensional modulated lattices, the investigation of the impact of interactions on the transport dynamics, or the use of optimal control to optimize the actual shape of the transporting state, in order \textit{e.g.} to maximize its initial overlap with the ground state of the potential. 


\paragraph*{Aknowledgements}
We thank Gabriel Chatelain and Maxime Martinez for helpful discussions as well as Alexandre Dugelay for experimental support. This work was (partially) supported through the EUR Grant NanoX No. ANR-17-EURE-0009 in the framework of the ``Programme d’Investissements d’Avenir'' and research funding Grants No. ANR-17-CE30-0024 and ANR-22-CE47-0008. N.D. and F.A. acknowledge support from R\'egion Occitanie and Universit\'e Toulouse III-Paul Sabatier. N.D. acknowledges ERC Grant LATIS for support in Brussels.

\bibliography{article_ratchet_bib}

\begin{thebibliography}{52}%
\makeatletter
\providecommand \@ifxundefined [1]{%
 \@ifx{#1\undefined}
}%
\providecommand \@ifnum [1]{%
 \ifnum #1\expandafter \@firstoftwo
 \else \expandafter \@secondoftwo
 \fi
}%
\providecommand \@ifx [1]{%
 \ifx #1\expandafter \@firstoftwo
 \else \expandafter \@secondoftwo
 \fi
}%
\providecommand \natexlab [1]{#1}%
\providecommand \enquote  [1]{``#1''}%
\providecommand \bibnamefont  [1]{#1}%
\providecommand \bibfnamefont [1]{#1}%
\providecommand \citenamefont [1]{#1}%
\providecommand \href@noop [0]{\@secondoftwo}%
\providecommand \href [0]{\begingroup \@sanitize@url \@href}%
\providecommand \@href[1]{\@@startlink{#1}\@@href}%
\providecommand \@@href[1]{\endgroup#1\@@endlink}%
\providecommand \@sanitize@url [0]{\catcode `\\12\catcode `\$12\catcode
  `\&12\catcode `\#12\catcode `\^12\catcode `\_12\catcode `\%12\relax}%
\providecommand \@@startlink[1]{}%
\providecommand \@@endlink[0]{}%
\providecommand \url  [0]{\begingroup\@sanitize@url \@url }%
\providecommand \@url [1]{\endgroup\@href {#1}{\urlprefix }}%
\providecommand \urlprefix  [0]{URL }%
\providecommand \Eprint [0]{\href }%
\providecommand \doibase [0]{https://doi.org/}%
\providecommand \selectlanguage [0]{\@gobble}%
\providecommand \bibinfo  [0]{\@secondoftwo}%
\providecommand \bibfield  [0]{\@secondoftwo}%
\providecommand \translation [1]{[#1]}%
\providecommand \BibitemOpen [0]{}%
\providecommand \bibitemStop [0]{}%
\providecommand \bibitemNoStop [0]{.\EOS\space}%
\providecommand \EOS [0]{\spacefactor3000\relax}%
\providecommand \BibitemShut  [1]{\csname bibitem#1\endcsname}%
\let\auto@bib@innerbib\@empty
\bibitem [{\citenamefont {Flach}\ \emph {et~al.}(2000)\citenamefont {Flach},
  \citenamefont {Yevtushenko},\ and\ \citenamefont {Zolotaryuk}}]{flach_2000}%
  \BibitemOpen
  \bibfield  {author} {\bibinfo {author} {\bibfnamefont {S.}~\bibnamefont
  {Flach}}, \bibinfo {author} {\bibfnamefont {O.}~\bibnamefont {Yevtushenko}},\
  and\ \bibinfo {author} {\bibfnamefont {Y.}~\bibnamefont {Zolotaryuk}},\
  }\bibfield  {title} {\bibinfo {title} {{Directed Current due to Broken
  Time-Space Symmetry}},\ }\href {https://doi.org/10.1103/PhysRevLett.84.2358}
  {\bibfield  {journal} {\bibinfo  {journal} {Phys. Rev. Lett.}\ }\textbf
  {\bibinfo {volume} {84}},\ \bibinfo {pages} {2358} (\bibinfo {year}
  {2000})}\BibitemShut {NoStop}%
\bibitem [{\citenamefont {Denisov}\ \emph {et~al.}(2002)\citenamefont
  {Denisov}, \citenamefont {Flach}, \citenamefont {Ovchinnikov}, \citenamefont
  {Yevtushenko},\ and\ \citenamefont {Zolotaryuk}}]{denisov_2002}%
  \BibitemOpen
  \bibfield  {author} {\bibinfo {author} {\bibfnamefont {S.}~\bibnamefont
  {Denisov}}, \bibinfo {author} {\bibfnamefont {S.}~\bibnamefont {Flach}},
  \bibinfo {author} {\bibfnamefont {A.~A.}\ \bibnamefont {Ovchinnikov}},
  \bibinfo {author} {\bibfnamefont {O.}~\bibnamefont {Yevtushenko}},\ and\
  \bibinfo {author} {\bibfnamefont {Y.}~\bibnamefont {Zolotaryuk}},\ }\bibfield
   {title} {\bibinfo {title} {Broken space-time symmetries and mechanisms of
  rectification of ac fields by nonlinear (non)adiabatic response},\ }\href
  {https://doi.org/10.1103/PhysRevE.66.041104} {\bibfield  {journal} {\bibinfo
  {journal} {Phys. Rev. E}\ }\textbf {\bibinfo {volume} {66}},\ \bibinfo
  {pages} {041104} (\bibinfo {year} {2002})}\BibitemShut {NoStop}%
\bibitem [{\citenamefont {Denisov}\ \emph {et~al.}(2014)\citenamefont
  {Denisov}, \citenamefont {Flach},\ and\ \citenamefont
  {Hänggi}}]{denisov_2014}%
  \BibitemOpen
  \bibfield  {author} {\bibinfo {author} {\bibfnamefont {S.}~\bibnamefont
  {Denisov}}, \bibinfo {author} {\bibfnamefont {S.}~\bibnamefont {Flach}},\
  and\ \bibinfo {author} {\bibfnamefont {P.}~\bibnamefont {Hänggi}},\
  }\bibfield  {title} {\bibinfo {title} {Tunable transport with broken
  space–time symmetries},\ }\href
  {https://doi.org/https://doi.org/10.1016/j.physrep.2014.01.003} {\bibfield
  {journal} {\bibinfo  {journal} {Physics Reports}\ }\textbf {\bibinfo {volume}
  {538}},\ \bibinfo {pages} {77} (\bibinfo {year} {2014})}\BibitemShut
  {NoStop}%
\bibitem [{\citenamefont {H\"anggi}\ and\ \citenamefont
  {Marchesoni}(2005)}]{hanggi_2005}%
  \BibitemOpen
  \bibfield  {author} {\bibinfo {author} {\bibfnamefont {P.}~\bibnamefont
  {H\"anggi}}\ and\ \bibinfo {author} {\bibfnamefont {F.}~\bibnamefont
  {Marchesoni}},\ }\bibfield  {title} {\bibinfo {title} {{Introduction: 100
  years of Brownian motion}},\ }\href {https://doi.org/10.1063/1.1895505}
  {\bibfield  {journal} {\bibinfo  {journal} {Chaos: An Interdisciplinary
  Journal of Nonlinear Science}\ }\textbf {\bibinfo {volume} {15}},\ \bibinfo
  {pages} {026101} (\bibinfo {year} {2005})}\BibitemShut {NoStop}%
\bibitem [{\citenamefont {Rousselet}\ \emph {et~al.}(1994)\citenamefont
  {Rousselet}, \citenamefont {Salome}, \citenamefont {Ajdari},\ and\
  \citenamefont {Prost}}]{rousselet_1994}%
  \BibitemOpen
  \bibfield  {author} {\bibinfo {author} {\bibfnamefont {J.}~\bibnamefont
  {Rousselet}}, \bibinfo {author} {\bibfnamefont {L.}~\bibnamefont {Salome}},
  \bibinfo {author} {\bibfnamefont {A.}~\bibnamefont {Ajdari}},\ and\ \bibinfo
  {author} {\bibfnamefont {J.}~\bibnamefont {Prost}},\ }\bibfield  {title}
  {\bibinfo {title} {Directional motion of brownian particles induced by a
  periodic asymmetric potential},\ }\href {https://doi.org/10.1038/370446a0}
  {\bibfield  {journal} {\bibinfo  {journal} {Nature}\ }\textbf {\bibinfo
  {volume} {370}},\ \bibinfo {pages} {446} (\bibinfo {year}
  {1994})}\BibitemShut {NoStop}%
\bibitem [{\citenamefont {Mennerat-Robilliard}\ \emph
  {et~al.}(1999)\citenamefont {Mennerat-Robilliard}, \citenamefont {Lucas},
  \citenamefont {Guibal}, \citenamefont {Tabosa}, \citenamefont {Jurczak},
  \citenamefont {Courtois},\ and\ \citenamefont {Grynberg}}]{mr_1999}%
  \BibitemOpen
  \bibfield  {author} {\bibinfo {author} {\bibfnamefont {C.}~\bibnamefont
  {Mennerat-Robilliard}}, \bibinfo {author} {\bibfnamefont {D.}~\bibnamefont
  {Lucas}}, \bibinfo {author} {\bibfnamefont {S.}~\bibnamefont {Guibal}},
  \bibinfo {author} {\bibfnamefont {J.}~\bibnamefont {Tabosa}}, \bibinfo
  {author} {\bibfnamefont {C.}~\bibnamefont {Jurczak}}, \bibinfo {author}
  {\bibfnamefont {J.-Y.}\ \bibnamefont {Courtois}},\ and\ \bibinfo {author}
  {\bibfnamefont {G.}~\bibnamefont {Grynberg}},\ }\bibfield  {title} {\bibinfo
  {title} {{Ratchet for Cold Rubidium Atoms: The Asymmetric Optical Lattice}},\
  }\href {https://doi.org/10.1103/PhysRevLett.82.851} {\bibfield  {journal}
  {\bibinfo  {journal} {Phys. Rev. Lett.}\ }\textbf {\bibinfo {volume} {82}},\
  \bibinfo {pages} {851} (\bibinfo {year} {1999})}\BibitemShut {NoStop}%
\bibitem [{\citenamefont {Astumian}\ and\ \citenamefont
  {H\"anggi}(2002)}]{astumian_2002}%
  \BibitemOpen
  \bibfield  {author} {\bibinfo {author} {\bibfnamefont {R.~D.}\ \bibnamefont
  {Astumian}}\ and\ \bibinfo {author} {\bibfnamefont {P.}~\bibnamefont
  {H\"anggi}},\ }\bibfield  {title} {\bibinfo {title} {Brownian motors},\
  }\href {https://doi.org/10.1063/1.1535005} {\bibfield  {journal} {\bibinfo
  {journal} {Physics Today}\ }\textbf {\bibinfo {volume} {55}},\ \bibinfo
  {pages} {33} (\bibinfo {year} {2002})}\BibitemShut {NoStop}%
\bibitem [{\citenamefont {Schiavoni}\ \emph {et~al.}(2003)\citenamefont
  {Schiavoni}, \citenamefont {Sanchez-Palencia}, \citenamefont {Renzoni},\ and\
  \citenamefont {Grynberg}}]{schiavoni_2003}%
  \BibitemOpen
  \bibfield  {author} {\bibinfo {author} {\bibfnamefont {M.}~\bibnamefont
  {Schiavoni}}, \bibinfo {author} {\bibfnamefont {L.}~\bibnamefont
  {Sanchez-Palencia}}, \bibinfo {author} {\bibfnamefont {F.}~\bibnamefont
  {Renzoni}},\ and\ \bibinfo {author} {\bibfnamefont {G.}~\bibnamefont
  {Grynberg}},\ }\bibfield  {title} {\bibinfo {title} {{Phase Control of
  Directed Diffusion in a Symmetric Optical Lattice}},\ }\href
  {https://doi.org/10.1103/PhysRevLett.90.094101} {\bibfield  {journal}
  {\bibinfo  {journal} {Phys. Rev. Lett.}\ }\textbf {\bibinfo {volume} {90}},\
  \bibinfo {pages} {094101} (\bibinfo {year} {2003})}\BibitemShut {NoStop}%
\bibitem [{\citenamefont {Sanchez-Palencia}(2004)}]{sanchez_2004}%
  \BibitemOpen
  \bibfield  {author} {\bibinfo {author} {\bibfnamefont {L.}~\bibnamefont
  {Sanchez-Palencia}},\ }\bibfield  {title} {\bibinfo {title} {{Directed
  transport of Brownian particles in a double symmetric potential}},\ }\href
  {https://doi.org/10.1103/PhysRevE.70.011102} {\bibfield  {journal} {\bibinfo
  {journal} {Phys. Rev. E}\ }\textbf {\bibinfo {volume} {70}},\ \bibinfo
  {pages} {011102} (\bibinfo {year} {2004})}\BibitemShut {NoStop}%
\bibitem [{\citenamefont {Sj\"olund}\ \emph {et~al.}(2006)\citenamefont
  {Sj\"olund}, \citenamefont {Petra}, \citenamefont {Dion}, \citenamefont
  {Jonsell}, \citenamefont {Nyl\'en}, \citenamefont {Sanchez-Palencia},\ and\
  \citenamefont {Kastberg}}]{sjolund_2006}%
  \BibitemOpen
  \bibfield  {author} {\bibinfo {author} {\bibfnamefont {P.}~\bibnamefont
  {Sj\"olund}}, \bibinfo {author} {\bibfnamefont {S.~J.~H.}\ \bibnamefont
  {Petra}}, \bibinfo {author} {\bibfnamefont {C.~M.}\ \bibnamefont {Dion}},
  \bibinfo {author} {\bibfnamefont {S.}~\bibnamefont {Jonsell}}, \bibinfo
  {author} {\bibfnamefont {M.}~\bibnamefont {Nyl\'en}}, \bibinfo {author}
  {\bibfnamefont {L.}~\bibnamefont {Sanchez-Palencia}},\ and\ \bibinfo {author}
  {\bibfnamefont {A.}~\bibnamefont {Kastberg}},\ }\bibfield  {title} {\bibinfo
  {title} {{Demonstration of a Controllable Three-Dimensional Brownian Motor in
  Symmetric Potentials}},\ }\href
  {https://doi.org/10.1103/PhysRevLett.96.190602} {\bibfield  {journal}
  {\bibinfo  {journal} {Phys. Rev. Lett.}\ }\textbf {\bibinfo {volume} {96}},\
  \bibinfo {pages} {190602} (\bibinfo {year} {2006})}\BibitemShut {NoStop}%
\bibitem [{\citenamefont {H\"anggi}\ and\ \citenamefont
  {Marchesoni}(2009)}]{hanggi_2009}%
  \BibitemOpen
  \bibfield  {author} {\bibinfo {author} {\bibfnamefont {P.}~\bibnamefont
  {H\"anggi}}\ and\ \bibinfo {author} {\bibfnamefont {F.}~\bibnamefont
  {Marchesoni}},\ }\bibfield  {title} {\bibinfo {title} {{Artificial Brownian
  motors: Controlling transport on the nanoscale}},\ }\href
  {https://doi.org/10.1103/RevModPhys.81.387} {\bibfield  {journal} {\bibinfo
  {journal} {Rev. Mod. Phys.}\ }\textbf {\bibinfo {volume} {81}},\ \bibinfo
  {pages} {387} (\bibinfo {year} {2009})}\BibitemShut {NoStop}%
\bibitem [{\citenamefont {J\"ulicher}\ \emph {et~al.}(1997)\citenamefont
  {J\"ulicher}, \citenamefont {Ajdari},\ and\ \citenamefont
  {Prost}}]{julicher_1997}%
  \BibitemOpen
  \bibfield  {author} {\bibinfo {author} {\bibfnamefont {F.}~\bibnamefont
  {J\"ulicher}}, \bibinfo {author} {\bibfnamefont {A.}~\bibnamefont {Ajdari}},\
  and\ \bibinfo {author} {\bibfnamefont {J.}~\bibnamefont {Prost}},\ }\bibfield
   {title} {\bibinfo {title} {Modeling molecular motors},\ }\href
  {https://doi.org/10.1103/RevModPhys.69.1269} {\bibfield  {journal} {\bibinfo
  {journal} {Rev. Mod. Phys.}\ }\textbf {\bibinfo {volume} {69}},\ \bibinfo
  {pages} {1269} (\bibinfo {year} {1997})}\BibitemShut {NoStop}%
\bibitem [{\citenamefont {Chowdhury}\ \emph {et~al.}(2007)\citenamefont
  {Chowdhury}, \citenamefont {Schadschneider},\ and\ \citenamefont
  {Nishinari}}]{chowdhury_2005}%
  \BibitemOpen
  \bibfield  {author} {\bibinfo {author} {\bibfnamefont {D.}~\bibnamefont
  {Chowdhury}}, \bibinfo {author} {\bibfnamefont {A.}~\bibnamefont
  {Schadschneider}},\ and\ \bibinfo {author} {\bibfnamefont {K.}~\bibnamefont
  {Nishinari}},\ }\bibfield  {title} {\bibinfo {title} {Traffic phenomena in
  biology: From molecular motors to organisms},\ }in\ \href
  {https://doi.org/10.1007/978-3-540-47641-2_18} {\emph {\bibinfo {booktitle}
  {Traffic and Granular Flow'05}}},\ \bibinfo {editor} {edited by\ \bibinfo
  {editor} {\bibfnamefont {A.}~\bibnamefont {Schadschneider}}, \bibinfo
  {editor} {\bibfnamefont {T.}~\bibnamefont {P{\"o}schel}}, \bibinfo {editor}
  {\bibfnamefont {R.}~\bibnamefont {K{\"u}hne}}, \bibinfo {editor}
  {\bibfnamefont {M.}~\bibnamefont {Schreckenberg}},\ and\ \bibinfo {editor}
  {\bibfnamefont {D.~E.}\ \bibnamefont {Wolf}}}\ (\bibinfo  {publisher}
  {Springer Berlin Heidelberg},\ \bibinfo {address} {Berlin, Heidelberg},\
  \bibinfo {year} {2007})\ pp.\ \bibinfo {pages} {223--238}\BibitemShut
  {NoStop}%
\bibitem [{\citenamefont {Nishiyama}\ \emph {et~al.}(2002)\citenamefont
  {Nishiyama}, \citenamefont {Higuchi},\ and\ \citenamefont
  {Yanagida}}]{nishiyama_2002}%
  \BibitemOpen
  \bibfield  {author} {\bibinfo {author} {\bibfnamefont {M.}~\bibnamefont
  {Nishiyama}}, \bibinfo {author} {\bibfnamefont {H.}~\bibnamefont {Higuchi}},\
  and\ \bibinfo {author} {\bibfnamefont {T.}~\bibnamefont {Yanagida}},\
  }\bibfield  {title} {\bibinfo {title} {Chemomechanical coupling of the
  forward and backward steps of single kinesin molecules},\ }\href
  {https://doi.org/10.1038/ncb857} {\bibfield  {journal} {\bibinfo  {journal}
  {Nature Cell Biology}\ }\textbf {\bibinfo {volume} {4}},\ \bibinfo {pages}
  {790} (\bibinfo {year} {2002})}\BibitemShut {NoStop}%
\bibitem [{\citenamefont {Reimann}(2002)}]{reimann_2002}%
  \BibitemOpen
  \bibfield  {author} {\bibinfo {author} {\bibfnamefont {P.}~\bibnamefont
  {Reimann}},\ }\bibfield  {title} {\bibinfo {title} {Brownian motors: noisy
  transport far from equilibrium},\ }\href
  {https://doi.org/https://doi.org/10.1016/S0370-1573(01)00081-3} {\bibfield
  {journal} {\bibinfo  {journal} {Physics Reports}\ }\textbf {\bibinfo {volume}
  {361}},\ \bibinfo {pages} {57} (\bibinfo {year} {2002})}\BibitemShut
  {NoStop}%
\bibitem [{\citenamefont {Jung}\ \emph {et~al.}(1996)\citenamefont {Jung},
  \citenamefont {Kissner},\ and\ \citenamefont {H\"anggi}}]{jung_1996}%
  \BibitemOpen
  \bibfield  {author} {\bibinfo {author} {\bibfnamefont {P.}~\bibnamefont
  {Jung}}, \bibinfo {author} {\bibfnamefont {J.~G.}\ \bibnamefont {Kissner}},\
  and\ \bibinfo {author} {\bibfnamefont {P.}~\bibnamefont {H\"anggi}},\
  }\bibfield  {title} {\bibinfo {title} {{Regular and Chaotic Transport in
  Asymmetric Periodic Potentials: Inertia Ratchets}},\ }\href
  {https://doi.org/10.1103/PhysRevLett.76.3436} {\bibfield  {journal} {\bibinfo
   {journal} {Phys. Rev. Lett.}\ }\textbf {\bibinfo {volume} {76}},\ \bibinfo
  {pages} {3436} (\bibinfo {year} {1996})}\BibitemShut {NoStop}%
\bibitem [{\citenamefont {Carlo}\ \emph {et~al.}(2005)\citenamefont {Carlo},
  \citenamefont {Benenti}, \citenamefont {Casati},\ and\ \citenamefont
  {Shepelyansky}}]{carlo_2005}%
  \BibitemOpen
  \bibfield  {author} {\bibinfo {author} {\bibfnamefont {G.~G.}\ \bibnamefont
  {Carlo}}, \bibinfo {author} {\bibfnamefont {G.}~\bibnamefont {Benenti}},
  \bibinfo {author} {\bibfnamefont {G.}~\bibnamefont {Casati}},\ and\ \bibinfo
  {author} {\bibfnamefont {D.~L.}\ \bibnamefont {Shepelyansky}},\ }\bibfield
  {title} {\bibinfo {title} {{Quantum Ratchets in Dissipative Chaotic
  Systems}},\ }\href {https://doi.org/10.1103/PhysRevLett.94.164101} {\bibfield
   {journal} {\bibinfo  {journal} {Phys. Rev. Lett.}\ }\textbf {\bibinfo
  {volume} {94}},\ \bibinfo {pages} {164101} (\bibinfo {year}
  {2005})}\BibitemShut {NoStop}%
\bibitem [{\citenamefont {Wang}\ \emph {et~al.}(2007)\citenamefont {Wang},
  \citenamefont {Benenti}, \citenamefont {Casati},\ and\ \citenamefont
  {Li}}]{wang_2007}%
  \BibitemOpen
  \bibfield  {author} {\bibinfo {author} {\bibfnamefont {L.}~\bibnamefont
  {Wang}}, \bibinfo {author} {\bibfnamefont {G.}~\bibnamefont {Benenti}},
  \bibinfo {author} {\bibfnamefont {G.}~\bibnamefont {Casati}},\ and\ \bibinfo
  {author} {\bibfnamefont {B.}~\bibnamefont {Li}},\ }\bibfield  {title}
  {\bibinfo {title} {{Ratchet Effect and the Transporting Islands in the
  Chaotic Sea}},\ }\href {https://doi.org/10.1103/PhysRevLett.99.244101}
  {\bibfield  {journal} {\bibinfo  {journal} {Phys. Rev. Lett.}\ }\textbf
  {\bibinfo {volume} {99}},\ \bibinfo {pages} {244101} (\bibinfo {year}
  {2007})}\BibitemShut {NoStop}%
\bibitem [{\citenamefont {Denisov}\ \emph {et~al.}(2009)\citenamefont
  {Denisov}, \citenamefont {Kohler},\ and\ \citenamefont
  {Hänggi}}]{denisov_2009}%
  \BibitemOpen
  \bibfield  {author} {\bibinfo {author} {\bibfnamefont {S.}~\bibnamefont
  {Denisov}}, \bibinfo {author} {\bibfnamefont {S.}~\bibnamefont {Kohler}},\
  and\ \bibinfo {author} {\bibfnamefont {P.}~\bibnamefont {Hänggi}},\
  }\bibfield  {title} {\bibinfo {title} {Underdamped quantum ratchets},\ }\href
  {https://doi.org/10.1209/0295-5075/85/40003} {\bibfield  {journal} {\bibinfo
  {journal} {{EPL} (Europhysics Letters)}\ }\textbf {\bibinfo {volume} {85}},\
  \bibinfo {pages} {40003} (\bibinfo {year} {2009})}\BibitemShut {NoStop}%
\bibitem [{\citenamefont {Schanz}\ \emph {et~al.}(2001)\citenamefont {Schanz},
  \citenamefont {Otto}, \citenamefont {Ketzmerick},\ and\ \citenamefont
  {Dittrich}}]{schanz_2001}%
  \BibitemOpen
  \bibfield  {author} {\bibinfo {author} {\bibfnamefont {H.}~\bibnamefont
  {Schanz}}, \bibinfo {author} {\bibfnamefont {M.-F.}\ \bibnamefont {Otto}},
  \bibinfo {author} {\bibfnamefont {R.}~\bibnamefont {Ketzmerick}},\ and\
  \bibinfo {author} {\bibfnamefont {T.}~\bibnamefont {Dittrich}},\ }\bibfield
  {title} {\bibinfo {title} {{Classical and Quantum Hamiltonian Ratchets}},\
  }\href {https://doi.org/10.1103/PhysRevLett.87.070601} {\bibfield  {journal}
  {\bibinfo  {journal} {Phys. Rev. Lett.}\ }\textbf {\bibinfo {volume} {87}},\
  \bibinfo {pages} {070601} (\bibinfo {year} {2001})}\BibitemShut {NoStop}%
\bibitem [{\citenamefont {Denisov}\ \emph {et~al.}(2007)\citenamefont
  {Denisov}, \citenamefont {Morales-Molina}, \citenamefont {Flach},\ and\
  \citenamefont {H\"anggi}}]{denisov_2007}%
  \BibitemOpen
  \bibfield  {author} {\bibinfo {author} {\bibfnamefont {S.}~\bibnamefont
  {Denisov}}, \bibinfo {author} {\bibfnamefont {L.}~\bibnamefont
  {Morales-Molina}}, \bibinfo {author} {\bibfnamefont {S.}~\bibnamefont
  {Flach}},\ and\ \bibinfo {author} {\bibfnamefont {P.}~\bibnamefont
  {H\"anggi}},\ }\bibfield  {title} {\bibinfo {title} {{Periodically driven
  quantum ratchets: Symmetries and resonances}},\ }\href
  {https://doi.org/10.1103/PhysRevA.75.063424} {\bibfield  {journal} {\bibinfo
  {journal} {Phys. Rev. A}\ }\textbf {\bibinfo {volume} {75}},\ \bibinfo
  {pages} {063424} (\bibinfo {year} {2007})}\BibitemShut {NoStop}%
\bibitem [{\citenamefont {Salger}\ \emph {et~al.}(2009)\citenamefont {Salger},
  \citenamefont {Kling}, \citenamefont {Hecking}, \citenamefont {Geckeler},
  \citenamefont {Morales-Molina},\ and\ \citenamefont {Weitz}}]{salger_2009}%
  \BibitemOpen
  \bibfield  {author} {\bibinfo {author} {\bibfnamefont {T.}~\bibnamefont
  {Salger}}, \bibinfo {author} {\bibfnamefont {S.}~\bibnamefont {Kling}},
  \bibinfo {author} {\bibfnamefont {T.}~\bibnamefont {Hecking}}, \bibinfo
  {author} {\bibfnamefont {C.}~\bibnamefont {Geckeler}}, \bibinfo {author}
  {\bibfnamefont {L.}~\bibnamefont {Morales-Molina}},\ and\ \bibinfo {author}
  {\bibfnamefont {M.}~\bibnamefont {Weitz}},\ }\bibfield  {title} {\bibinfo
  {title} {{Directed Transport of Atoms in a Hamiltonian Quantum Ratchet}},\
  }\href {https://doi.org/10.1126/science.1179546} {\bibfield  {journal}
  {\bibinfo  {journal} {Science}\ }\textbf {\bibinfo {volume} {326}},\ \bibinfo
  {pages} {1241} (\bibinfo {year} {2009})}\BibitemShut {NoStop}%
\bibitem [{\citenamefont {Zhan}\ \emph {et~al.}(2011)\citenamefont {Zhan},
  \citenamefont {Denisov}, \citenamefont {Ponomarev},\ and\ \citenamefont
  {Hänggi}}]{zhan_2011}%
  \BibitemOpen
  \bibfield  {author} {\bibinfo {author} {\bibfnamefont {F.}~\bibnamefont
  {Zhan}}, \bibinfo {author} {\bibfnamefont {S.}~\bibnamefont {Denisov}},
  \bibinfo {author} {\bibfnamefont {A.~V.}\ \bibnamefont {Ponomarev}},\ and\
  \bibinfo {author} {\bibfnamefont {P.}~\bibnamefont {Hänggi}},\ }\bibfield
  {title} {\bibinfo {title} {Quantum ratchet transport with minimal dispersion
  rate},\ }\href {https://doi.org/10.1103/PhysRevA.84.043617} {\bibfield
  {journal} {\bibinfo  {journal} {Physical Review A}\ }\textbf {\bibinfo
  {volume} {84}},\ \bibinfo {pages} {043617} (\bibinfo {year}
  {2011})}\BibitemShut {NoStop}%
\bibitem [{\citenamefont {Lundh}\ and\ \citenamefont
  {Wallin}(2005)}]{lundh_2005}%
  \BibitemOpen
  \bibfield  {author} {\bibinfo {author} {\bibfnamefont {E.}~\bibnamefont
  {Lundh}}\ and\ \bibinfo {author} {\bibfnamefont {M.}~\bibnamefont {Wallin}},\
  }\bibfield  {title} {\bibinfo {title} {Ratchet effect for cold atoms in an
  optical lattice},\ }\href {https://doi.org/10.1103/PhysRevLett.94.110603}
  {\bibfield  {journal} {\bibinfo  {journal} {Phys. Rev. Lett.}\ }\textbf
  {\bibinfo {volume} {94}},\ \bibinfo {pages} {110603} (\bibinfo {year}
  {2005})}\BibitemShut {NoStop}%
\bibitem [{\citenamefont {Ni}\ \emph {et~al.}(2017)\citenamefont {Ni},
  \citenamefont {Dadras}, \citenamefont {Lam}, \citenamefont {Shrestha},
  \citenamefont {Sadgrove}, \citenamefont {Wimberger},\ and\ \citenamefont
  {Summy}}]{ni_2017}%
  \BibitemOpen
  \bibfield  {author} {\bibinfo {author} {\bibfnamefont {J.}~\bibnamefont
  {Ni}}, \bibinfo {author} {\bibfnamefont {S.}~\bibnamefont {Dadras}}, \bibinfo
  {author} {\bibfnamefont {W.~K.}\ \bibnamefont {Lam}}, \bibinfo {author}
  {\bibfnamefont {R.~K.}\ \bibnamefont {Shrestha}}, \bibinfo {author}
  {\bibfnamefont {M.}~\bibnamefont {Sadgrove}}, \bibinfo {author}
  {\bibfnamefont {S.}~\bibnamefont {Wimberger}},\ and\ \bibinfo {author}
  {\bibfnamefont {G.~S.}\ \bibnamefont {Summy}},\ }\bibfield  {title} {\bibinfo
  {title} {{Hamiltonian Ratchets with Ultra-Cold Atoms}},\ }\href
  {https://doi.org/https://doi.org/10.1002/andp.201600335} {\bibfield
  {journal} {\bibinfo  {journal} {Annalen der Physik}\ }\textbf {\bibinfo
  {volume} {529}},\ \bibinfo {pages} {1600335} (\bibinfo {year}
  {2017})}\BibitemShut {NoStop}%
\bibitem [{\citenamefont {Citro}\ and\ \citenamefont
  {Aidelsburger}(2023)}]{Citro_2023}%
  \BibitemOpen
  \bibfield  {author} {\bibinfo {author} {\bibfnamefont {R.}~\bibnamefont
  {Citro}}\ and\ \bibinfo {author} {\bibfnamefont {M.}~\bibnamefont
  {Aidelsburger}},\ }\bibfield  {title} {\bibinfo {title} {Thouless pumping and
  topology},\ }\href {https://doi.org/10.1038/s42254-022-00545-0} {\bibfield
  {journal} {\bibinfo  {journal} {Nature Reviews Physics}\ }\textbf {\bibinfo
  {volume} {5}},\ \bibinfo {pages} {87} (\bibinfo {year} {2023})}\BibitemShut
  {NoStop}%
\bibitem [{\citenamefont {Gong}\ and\ \citenamefont
  {Brumer}(2004)}]{gong_2004}%
  \BibitemOpen
  \bibfield  {author} {\bibinfo {author} {\bibfnamefont {J.}~\bibnamefont
  {Gong}}\ and\ \bibinfo {author} {\bibfnamefont {P.}~\bibnamefont {Brumer}},\
  }\bibfield  {title} {\bibinfo {title} {Directed anomalous diffusion without a
  biased field: A ratchet accelerator},\ }\href
  {https://doi.org/10.1103/PhysRevE.70.016202} {\bibfield  {journal} {\bibinfo
  {journal} {Phys. Rev. E}\ }\textbf {\bibinfo {volume} {70}},\ \bibinfo
  {pages} {016202} (\bibinfo {year} {2004})}\BibitemShut {NoStop}%
\bibitem [{\citenamefont {Sadgrove}\ \emph {et~al.}(2013)\citenamefont
  {Sadgrove}, \citenamefont {Schell}, \citenamefont {Nakagawa},\ and\
  \citenamefont {Wimberger}}]{sadgrove_2013}%
  \BibitemOpen
  \bibfield  {author} {\bibinfo {author} {\bibfnamefont {M.}~\bibnamefont
  {Sadgrove}}, \bibinfo {author} {\bibfnamefont {T.}~\bibnamefont {Schell}},
  \bibinfo {author} {\bibfnamefont {K.}~\bibnamefont {Nakagawa}},\ and\
  \bibinfo {author} {\bibfnamefont {S.}~\bibnamefont {Wimberger}},\ }\bibfield
  {title} {\bibinfo {title} {Engineering quantum correlations to enhance
  transport in cold atoms},\ }\href
  {https://doi.org/10.1103/PhysRevA.87.013631} {\bibfield  {journal} {\bibinfo
  {journal} {Phys. Rev. A}\ }\textbf {\bibinfo {volume} {87}},\ \bibinfo
  {pages} {013631} (\bibinfo {year} {2013})}\BibitemShut {NoStop}%
\bibitem [{\citenamefont {White}\ \emph {et~al.}(2013)\citenamefont {White},
  \citenamefont {Ruddell},\ and\ \citenamefont {Hoogerland}}]{white_2013}%
  \BibitemOpen
  \bibfield  {author} {\bibinfo {author} {\bibfnamefont {D.~H.}\ \bibnamefont
  {White}}, \bibinfo {author} {\bibfnamefont {S.~K.}\ \bibnamefont {Ruddell}},\
  and\ \bibinfo {author} {\bibfnamefont {M.~D.}\ \bibnamefont {Hoogerland}},\
  }\bibfield  {title} {\bibinfo {title} {Experimental realization of a quantum
  ratchet through phase modulation},\ }\href
  {https://doi.org/10.1103/PhysRevA.88.063603} {\bibfield  {journal} {\bibinfo
  {journal} {Phys. Rev. A}\ }\textbf {\bibinfo {volume} {88}},\ \bibinfo
  {pages} {063603} (\bibinfo {year} {2013})}\BibitemShut {NoStop}%
\bibitem [{\citenamefont {Hainaut}\ \emph {et~al.}(2018)\citenamefont
  {Hainaut}, \citenamefont {Ran\ifmmode~\mbox{\c{c}}\else \c{c}\fi{}on},
  \citenamefont {Cl\'ement}, \citenamefont {Garreau}, \citenamefont
  {Szriftgiser}, \citenamefont {Chicireanu},\ and\ \citenamefont
  {Delande}}]{hainaut_2018}%
  \BibitemOpen
  \bibfield  {author} {\bibinfo {author} {\bibfnamefont {C.}~\bibnamefont
  {Hainaut}}, \bibinfo {author} {\bibfnamefont {A.}~\bibnamefont
  {Ran\ifmmode~\mbox{\c{c}}\else \c{c}\fi{}on}}, \bibinfo {author}
  {\bibfnamefont {J.-F.}\ \bibnamefont {Cl\'ement}}, \bibinfo {author}
  {\bibfnamefont {J.~C.}\ \bibnamefont {Garreau}}, \bibinfo {author}
  {\bibfnamefont {P.}~\bibnamefont {Szriftgiser}}, \bibinfo {author}
  {\bibfnamefont {R.}~\bibnamefont {Chicireanu}},\ and\ \bibinfo {author}
  {\bibfnamefont {D.}~\bibnamefont {Delande}},\ }\bibfield  {title} {\bibinfo
  {title} {Ratchet effect in the quantum kicked rotor and its destruction by
  dynamical localization},\ }\href {https://doi.org/10.1103/PhysRevA.97.061601}
  {\bibfield  {journal} {\bibinfo  {journal} {Phys. Rev. A}\ }\textbf {\bibinfo
  {volume} {97}},\ \bibinfo {pages} {061601} (\bibinfo {year}
  {2018})}\BibitemShut {NoStop}%
\bibitem [{\citenamefont {Dittrich}\ \emph {et~al.}(2000)\citenamefont
  {Dittrich}, \citenamefont {Ketzmerick}, \citenamefont {Otto},\ and\
  \citenamefont {Schanz}}]{dittrich_2000}%
  \BibitemOpen
  \bibfield  {author} {\bibinfo {author} {\bibfnamefont {T.}~\bibnamefont
  {Dittrich}}, \bibinfo {author} {\bibfnamefont {R.}~\bibnamefont
  {Ketzmerick}}, \bibinfo {author} {\bibfnamefont {M.-F.}\ \bibnamefont
  {Otto}},\ and\ \bibinfo {author} {\bibfnamefont {H.}~\bibnamefont {Schanz}},\
  }\bibfield  {title} {\bibinfo {title} {Classical and quantum transport in
  deterministic hamiltonian ratchets},\ }\href
  {https://doi.org/https://doi.org/10.1002/andp.200051209-1011} {\bibfield
  {journal} {\bibinfo  {journal} {Annalen der Physik}\ }\textbf {\bibinfo
  {volume} {512}},\ \bibinfo {pages} {755} (\bibinfo {year}
  {2000})}\BibitemShut {NoStop}%
\bibitem [{\citenamefont {Borromeo}\ and\ \citenamefont
  {Marchesoni}(2005)}]{borromeo_2005}%
  \BibitemOpen
  \bibfield  {author} {\bibinfo {author} {\bibfnamefont {M.}~\bibnamefont
  {Borromeo}}\ and\ \bibinfo {author} {\bibfnamefont {F.}~\bibnamefont
  {Marchesoni}},\ }\bibfield  {title} {\bibinfo {title} {Noise-assisted
  transport on symmetric periodic substrates},\ }\href
  {https://doi.org/10.1063/1.1858651} {\bibfield  {journal} {\bibinfo
  {journal} {Chaos: An Interdisciplinary Journal of Nonlinear Science}\
  }\textbf {\bibinfo {volume} {15}},\ \bibinfo {pages} {026110} (\bibinfo
  {year} {2005})}\BibitemShut {NoStop}%
\bibitem [{\citenamefont {Gommers}\ \emph {et~al.}(2008)\citenamefont
  {Gommers}, \citenamefont {Lebedev}, \citenamefont {Brown},\ and\
  \citenamefont {Renzoni}}]{gommers_2008}%
  \BibitemOpen
  \bibfield  {author} {\bibinfo {author} {\bibfnamefont {R.}~\bibnamefont
  {Gommers}}, \bibinfo {author} {\bibfnamefont {V.}~\bibnamefont {Lebedev}},
  \bibinfo {author} {\bibfnamefont {M.}~\bibnamefont {Brown}},\ and\ \bibinfo
  {author} {\bibfnamefont {F.}~\bibnamefont {Renzoni}},\ }\bibfield  {title}
  {\bibinfo {title} {{Gating Ratchet for Cold Atoms}},\ }\href
  {https://doi.org/10.1103/PhysRevLett.100.040603} {\bibfield  {journal}
  {\bibinfo  {journal} {Phys. Rev. Lett.}\ }\textbf {\bibinfo {volume} {100}},\
  \bibinfo {pages} {040603} (\bibinfo {year} {2008})}\BibitemShut {NoStop}%
\bibitem [{\citenamefont {Grossert}\ \emph {et~al.}(2016)\citenamefont
  {Grossert}, \citenamefont {Leder}, \citenamefont {Denisov}, \citenamefont
  {H{\"a}nggi},\ and\ \citenamefont {Weitz}}]{grossert_2016}%
  \BibitemOpen
  \bibfield  {author} {\bibinfo {author} {\bibfnamefont {C.}~\bibnamefont
  {Grossert}}, \bibinfo {author} {\bibfnamefont {M.}~\bibnamefont {Leder}},
  \bibinfo {author} {\bibfnamefont {S.}~\bibnamefont {Denisov}}, \bibinfo
  {author} {\bibfnamefont {P.}~\bibnamefont {H{\"a}nggi}},\ and\ \bibinfo
  {author} {\bibfnamefont {M.}~\bibnamefont {Weitz}},\ }\bibfield  {title}
  {\bibinfo {title} {Experimental control of transport resonances in a coherent
  quantum rocking ratchet},\ }\href {https://doi.org/10.1038/ncomms10440}
  {\bibfield  {journal} {\bibinfo  {journal} {Nature Communications}\ }\textbf
  {\bibinfo {volume} {7}},\ \bibinfo {pages} {10440} (\bibinfo {year}
  {2016})}\BibitemShut {NoStop}%
\bibitem [{\citenamefont {Tomsovic}\ and\ \citenamefont
  {Ullmo}(1994)}]{tomsovic_1994}%
  \BibitemOpen
  \bibfield  {author} {\bibinfo {author} {\bibfnamefont {S.}~\bibnamefont
  {Tomsovic}}\ and\ \bibinfo {author} {\bibfnamefont {D.}~\bibnamefont
  {Ullmo}},\ }\bibfield  {title} {\bibinfo {title} {Chaos-assisted tunneling},\
  }\href {https://doi.org/10.1103/PhysRevE.50.145} {\bibfield  {journal}
  {\bibinfo  {journal} {Phys. Rev. E}\ }\textbf {\bibinfo {volume} {50}},\
  \bibinfo {pages} {145} (\bibinfo {year} {1994})}\BibitemShut {NoStop}%
\bibitem [{\citenamefont {Arnal}\ \emph {et~al.}(2020)\citenamefont {Arnal},
  \citenamefont {Chatelain}, \citenamefont {Martinez}, \citenamefont {Dupont},
  \citenamefont {Giraud}, \citenamefont {Ullmo}, \citenamefont {Georgeot},
  \citenamefont {Lemari\'e}, \citenamefont {Billy},\ and\ \citenamefont
  {Gu\'ery-Odelin}}]{arnal_2020}%
  \BibitemOpen
  \bibfield  {author} {\bibinfo {author} {\bibfnamefont {M.}~\bibnamefont
  {Arnal}}, \bibinfo {author} {\bibfnamefont {G.}~\bibnamefont {Chatelain}},
  \bibinfo {author} {\bibfnamefont {M.}~\bibnamefont {Martinez}}, \bibinfo
  {author} {\bibfnamefont {N.}~\bibnamefont {Dupont}}, \bibinfo {author}
  {\bibfnamefont {O.}~\bibnamefont {Giraud}}, \bibinfo {author} {\bibfnamefont
  {D.}~\bibnamefont {Ullmo}}, \bibinfo {author} {\bibfnamefont
  {B.}~\bibnamefont {Georgeot}}, \bibinfo {author} {\bibfnamefont
  {G.}~\bibnamefont {Lemari\'e}}, \bibinfo {author} {\bibfnamefont
  {J.}~\bibnamefont {Billy}},\ and\ \bibinfo {author} {\bibfnamefont
  {D.}~\bibnamefont {Gu\'ery-Odelin}},\ }\bibfield  {title} {\bibinfo {title}
  {{Chaos-assisted tunneling resonances in a synthetic Floquet superlattice}},\
  }\href {https://doi.org/10.1126/sciadv.abc4886} {\bibfield  {journal}
  {\bibinfo  {journal} {Science Advances}\ }\textbf {\bibinfo {volume} {6}},\
  \bibinfo {pages} {eabc4886} (\bibinfo {year} {2020})}\BibitemShut {NoStop}%
\bibitem [{\citenamefont {Keshavamurthy}\ and\ \citenamefont
  {Schlagheck}(2011)}]{book_keshavamurthy_2011}%
  \BibitemOpen
  \bibinfo {editor} {\bibfnamefont {S.}~\bibnamefont {Keshavamurthy}}\ and\
  \bibinfo {editor} {\bibfnamefont {P.}~\bibnamefont {Schlagheck}},\ eds.,\
  \href
  {https://www.routledge.com/Dynamical-Tunneling-Theory-and-Experiment/Keshavamurthy-Schlagheck/p/book/9781138113503}
  {\emph {\bibinfo {title} {{Dynamical Tunneling: Theory and Experiment}}}}\
  (\bibinfo  {publisher} {CRC Press},\ \bibinfo {year} {2011})\BibitemShut
  {NoStop}%
\bibitem [{\citenamefont {Boscain}\ \emph {et~al.}(2021)\citenamefont
  {Boscain}, \citenamefont {Sigalotti},\ and\ \citenamefont
  {Sugny}}]{boscain_2021}%
  \BibitemOpen
  \bibfield  {author} {\bibinfo {author} {\bibfnamefont {U.}~\bibnamefont
  {Boscain}}, \bibinfo {author} {\bibfnamefont {M.}~\bibnamefont {Sigalotti}},\
  and\ \bibinfo {author} {\bibfnamefont {D.}~\bibnamefont {Sugny}},\ }\bibfield
   {title} {\bibinfo {title} {{Introduction to the Pontryagin Maximum Principle
  for Quantum Optimal Control}},\ }\href
  {https://doi.org/10.1103/PRXQuantum.2.030203} {\bibfield  {journal} {\bibinfo
   {journal} {PRX Quantum}\ }\textbf {\bibinfo {volume} {2}},\ \bibinfo {pages}
  {030203} (\bibinfo {year} {2021})}\BibitemShut {NoStop}%
\bibitem [{\citenamefont {Koch}\ \emph {et~al.}(2022)\citenamefont {Koch},
  \citenamefont {Boscain}, \citenamefont {Calarco}, \citenamefont {Dirr},
  \citenamefont {Filipp}, \citenamefont {Glaser}, \citenamefont {Kosloff},
  \citenamefont {Montangero}, \citenamefont {Schulte-Herbr{\"u}ggen},
  \citenamefont {Sugny},\ and\ \citenamefont {Wilhelm}}]{koch_2022}%
  \BibitemOpen
  \bibfield  {author} {\bibinfo {author} {\bibfnamefont {C.~P.}\ \bibnamefont
  {Koch}}, \bibinfo {author} {\bibfnamefont {U.}~\bibnamefont {Boscain}},
  \bibinfo {author} {\bibfnamefont {T.}~\bibnamefont {Calarco}}, \bibinfo
  {author} {\bibfnamefont {G.}~\bibnamefont {Dirr}}, \bibinfo {author}
  {\bibfnamefont {S.}~\bibnamefont {Filipp}}, \bibinfo {author} {\bibfnamefont
  {S.~J.}\ \bibnamefont {Glaser}}, \bibinfo {author} {\bibfnamefont
  {R.}~\bibnamefont {Kosloff}}, \bibinfo {author} {\bibfnamefont
  {S.}~\bibnamefont {Montangero}}, \bibinfo {author} {\bibfnamefont
  {T.}~\bibnamefont {Schulte-Herbr{\"u}ggen}}, \bibinfo {author} {\bibfnamefont
  {D.}~\bibnamefont {Sugny}},\ and\ \bibinfo {author} {\bibfnamefont {F.~K.}\
  \bibnamefont {Wilhelm}},\ }\bibfield  {title} {\bibinfo {title} {{Quantum
  optimal control in quantum technologies. Strategic report on current status,
  visions and goals for research in Europe}},\ }\href
  {https://doi.org/10.1140/epjqt/s40507-022-00138-x} {\bibfield  {journal}
  {\bibinfo  {journal} {EPJ Quantum Technology}\ }\textbf {\bibinfo {volume}
  {9}},\ \bibinfo {pages} {19} (\bibinfo {year} {2022})}\BibitemShut {NoStop}%
\bibitem [{\citenamefont {Dupont}(2022)}]{DupontPhD}%
  \BibitemOpen
  \bibfield  {author} {\bibinfo {author} {\bibfnamefont {N.}~\bibnamefont
  {Dupont}},\ }\emph {\bibinfo {title} {Control and transport of matter waves
  in an optical lattice}},\ \href {https://theses.hal.science/tel-03997401v1}
  {Ph.D. thesis},\ \bibinfo  {school} {Université Toulouse 3 - Paul Sabatier}
  (\bibinfo {year} {2022})\BibitemShut {NoStop}%
\bibitem [{\citenamefont {Berry}\ and\ \citenamefont
  {Robnik}(1984)}]{berry_1984}%
  \BibitemOpen
  \bibfield  {author} {\bibinfo {author} {\bibfnamefont {M.~V.}\ \bibnamefont
  {Berry}}\ and\ \bibinfo {author} {\bibfnamefont {M.}~\bibnamefont {Robnik}},\
  }\bibfield  {title} {\bibinfo {title} {Semiclassical level spacings when
  regular and chaotic orbits coexist},\ }\href
  {https://doi.org/10.1088/0305-4470/17/12/013} {\bibfield  {journal} {\bibinfo
   {journal} {Journal of Physics A: Mathematical and General}\ }\textbf
  {\bibinfo {volume} {17}},\ \bibinfo {pages} {2413} (\bibinfo {year}
  {1984})}\BibitemShut {NoStop}%
\bibitem [{\citenamefont {Bohigas}\ \emph {et~al.}(1993)\citenamefont
  {Bohigas}, \citenamefont {Tomsovic},\ and\ \citenamefont
  {Ullmo}}]{bohigas_1993}%
  \BibitemOpen
  \bibfield  {author} {\bibinfo {author} {\bibfnamefont {O.}~\bibnamefont
  {Bohigas}}, \bibinfo {author} {\bibfnamefont {S.}~\bibnamefont {Tomsovic}},\
  and\ \bibinfo {author} {\bibfnamefont {D.}~\bibnamefont {Ullmo}},\ }\bibfield
   {title} {\bibinfo {title} {Manifestations of classical phase space
  structures in quantum mechanics},\ }\href
  {https://doi.org/https://doi.org/10.1016/0370-1573(93)90109-Q} {\bibfield
  {journal} {\bibinfo  {journal} {Physics Reports}\ }\textbf {\bibinfo {volume}
  {223}},\ \bibinfo {pages} {43} (\bibinfo {year} {1993})}\BibitemShut
  {NoStop}%
\bibitem [{\citenamefont {Bahr}\ and\ \citenamefont
  {Korsch}(2007)}]{bahr_2007}%
  \BibitemOpen
  \bibfield  {author} {\bibinfo {author} {\bibfnamefont {B.}~\bibnamefont
  {Bahr}}\ and\ \bibinfo {author} {\bibfnamefont {H.~J.}\ \bibnamefont
  {Korsch}},\ }\bibfield  {title} {\bibinfo {title} {{Quantum mechanics on a
  circle: Husimi phase-space distributions and semiclassical coherent state
  propagators}},\ }\href {https://doi.org/10.1088/1751-8113/40/14/013}
  {\bibfield  {journal} {\bibinfo  {journal} {Journal of Physics A:
  Mathematical and Theoretical}\ }\textbf {\bibinfo {volume} {40}},\ \bibinfo
  {pages} {3959} (\bibinfo {year} {2007})}\BibitemShut {NoStop}%
\bibitem [{\citenamefont {Dupont}\ \emph {et~al.}(2023)\citenamefont {Dupont},
  \citenamefont {Arrouas}, \citenamefont {Gabardos}, \citenamefont {Ombredane},
  \citenamefont {Billy}, \citenamefont {Peaudecerf}, \citenamefont {Sugny},\
  and\ \citenamefont {Guéry-Odelin}}]{dupont_2023}%
  \BibitemOpen
  \bibfield  {author} {\bibinfo {author} {\bibfnamefont {N.}~\bibnamefont
  {Dupont}}, \bibinfo {author} {\bibfnamefont {F.}~\bibnamefont {Arrouas}},
  \bibinfo {author} {\bibfnamefont {L.}~\bibnamefont {Gabardos}}, \bibinfo
  {author} {\bibfnamefont {N.}~\bibnamefont {Ombredane}}, \bibinfo {author}
  {\bibfnamefont {J.}~\bibnamefont {Billy}}, \bibinfo {author} {\bibfnamefont
  {B.}~\bibnamefont {Peaudecerf}}, \bibinfo {author} {\bibfnamefont
  {D.}~\bibnamefont {Sugny}},\ and\ \bibinfo {author} {\bibfnamefont
  {D.}~\bibnamefont {Guéry-Odelin}},\ }\bibfield  {title} {\bibinfo {title}
  {{Phase-space distributions of Bose–Einstein condensates in an optical
  lattice: optimal shaping and reconstruction}},\ }\href
  {https://doi.org/10.1088/1367-2630/acaf9a} {\bibfield  {journal} {\bibinfo
  {journal} {New Journal of Physics}\ }\textbf {\bibinfo {volume} {25}},\
  \bibinfo {pages} {013012} (\bibinfo {year} {2023})}\BibitemShut {NoStop}%
\bibitem [{\citenamefont {Fortun}\ \emph {et~al.}(2016)\citenamefont {Fortun},
  \citenamefont {Cabrera-Guti\'errez}, \citenamefont {Condon}, \citenamefont
  {Michon}, \citenamefont {Billy},\ and\ \citenamefont
  {Gu\'ery-Odelin}}]{fortun_2016}%
  \BibitemOpen
  \bibfield  {author} {\bibinfo {author} {\bibfnamefont {A.}~\bibnamefont
  {Fortun}}, \bibinfo {author} {\bibfnamefont {C.}~\bibnamefont
  {Cabrera-Guti\'errez}}, \bibinfo {author} {\bibfnamefont {G.}~\bibnamefont
  {Condon}}, \bibinfo {author} {\bibfnamefont {E.}~\bibnamefont {Michon}},
  \bibinfo {author} {\bibfnamefont {J.}~\bibnamefont {Billy}},\ and\ \bibinfo
  {author} {\bibfnamefont {D.}~\bibnamefont {Gu\'ery-Odelin}},\ }\bibfield
  {title} {\bibinfo {title} {{Direct Tunneling Delay Time Measurement in an
  Optical Lattice}},\ }\href {https://doi.org/10.1103/PhysRevLett.117.010401}
  {\bibfield  {journal} {\bibinfo  {journal} {Phys. Rev. Lett.}\ }\textbf
  {\bibinfo {volume} {117}},\ \bibinfo {pages} {010401} (\bibinfo {year}
  {2016})}\BibitemShut {NoStop}%
\bibitem [{\citenamefont {Cabrera-Guti\'errez}\ \emph
  {et~al.}(2018)\citenamefont {Cabrera-Guti\'errez}, \citenamefont {Michon},
  \citenamefont {Brunaud}, \citenamefont {Kawalec}, \citenamefont {Fortun},
  \citenamefont {Arnal}, \citenamefont {Billy},\ and\ \citenamefont
  {Gu\'ery-Odelin}}]{cabrera_2018}%
  \BibitemOpen
  \bibfield  {author} {\bibinfo {author} {\bibfnamefont {C.}~\bibnamefont
  {Cabrera-Guti\'errez}}, \bibinfo {author} {\bibfnamefont {E.}~\bibnamefont
  {Michon}}, \bibinfo {author} {\bibfnamefont {V.}~\bibnamefont {Brunaud}},
  \bibinfo {author} {\bibfnamefont {T.}~\bibnamefont {Kawalec}}, \bibinfo
  {author} {\bibfnamefont {A.}~\bibnamefont {Fortun}}, \bibinfo {author}
  {\bibfnamefont {M.}~\bibnamefont {Arnal}}, \bibinfo {author} {\bibfnamefont
  {J.}~\bibnamefont {Billy}},\ and\ \bibinfo {author} {\bibfnamefont
  {D.}~\bibnamefont {Gu\'ery-Odelin}},\ }\bibfield  {title} {\bibinfo {title}
  {Robust calibration of an optical-lattice depth based on a phase shift},\
  }\href {https://doi.org/10.1103/PhysRevA.97.043617} {\bibfield  {journal}
  {\bibinfo  {journal} {Phys. Rev. A}\ }\textbf {\bibinfo {volume} {97}},\
  \bibinfo {pages} {043617} (\bibinfo {year} {2018})}\BibitemShut {NoStop}%
\bibitem [{Note1()}]{Note1}%
  \BibitemOpen
  \bibinfo {note} {Having $\gamma = s (E_\protect \text {L}/\hbar \omega )^2$
  and $\hbar _\protect \text {eff}= 4\pi ^2 \hbar / m\omega d^2$, $\hbar
  _\protect \text {eff}$ is varied without altering the classical dynamics by
  adjusting the lattice depth $s = 4\gamma / \hbar _\protect \text
  {eff}^2$.}\BibitemShut {Stop}%
\bibitem [{\citenamefont {Dupont}\ \emph {et~al.}(2021)\citenamefont {Dupont},
  \citenamefont {Chatelain}, \citenamefont {Gabardos}, \citenamefont {Arnal},
  \citenamefont {Billy}, \citenamefont {Peaudecerf}, \citenamefont {Sugny},\
  and\ \citenamefont {Gu\'ery-Odelin}}]{dupont_2021}%
  \BibitemOpen
  \bibfield  {author} {\bibinfo {author} {\bibfnamefont {N.}~\bibnamefont
  {Dupont}}, \bibinfo {author} {\bibfnamefont {G.}~\bibnamefont {Chatelain}},
  \bibinfo {author} {\bibfnamefont {L.}~\bibnamefont {Gabardos}}, \bibinfo
  {author} {\bibfnamefont {M.}~\bibnamefont {Arnal}}, \bibinfo {author}
  {\bibfnamefont {J.}~\bibnamefont {Billy}}, \bibinfo {author} {\bibfnamefont
  {B.}~\bibnamefont {Peaudecerf}}, \bibinfo {author} {\bibfnamefont
  {D.}~\bibnamefont {Sugny}},\ and\ \bibinfo {author} {\bibfnamefont
  {D.}~\bibnamefont {Gu\'ery-Odelin}},\ }\bibfield  {title} {\bibinfo {title}
  {{Quantum State Control of a Bose-Einstein Condensate in an Optical
  Lattice}},\ }\href {https://doi.org/10.1103/PRXQuantum.2.040303} {\bibfield
  {journal} {\bibinfo  {journal} {PRX Quantum}\ }\textbf {\bibinfo {volume}
  {2}},\ \bibinfo {pages} {040303} (\bibinfo {year} {2021})}\BibitemShut
  {NoStop}%
\bibitem [{Note2()}]{Note2}%
  \BibitemOpen
  \bibinfo {note} {This fidelity figure is set as a convergence condition on
  the iterative QOC algorithm~\cite {dupont_2021}.}\BibitemShut {Stop}%
\bibitem [{\citenamefont {Schrader}\ \emph {et~al.}(2001)\citenamefont
  {Schrader}, \citenamefont {Kuhr}, \citenamefont {Alt}, \citenamefont
  {Müller}, \citenamefont {Gomer},\ and\ \citenamefont
  {Meschede}}]{schrader_2001}%
  \BibitemOpen
  \bibfield  {author} {\bibinfo {author} {\bibfnamefont {D.}~\bibnamefont
  {Schrader}}, \bibinfo {author} {\bibfnamefont {S.}~\bibnamefont {Kuhr}},
  \bibinfo {author} {\bibfnamefont {W.}~\bibnamefont {Alt}}, \bibinfo {author}
  {\bibfnamefont {M.}~\bibnamefont {Müller}}, \bibinfo {author} {\bibfnamefont
  {V.}~\bibnamefont {Gomer}},\ and\ \bibinfo {author} {\bibfnamefont
  {D.}~\bibnamefont {Meschede}},\ }\bibfield  {title} {\bibinfo {title} {An
  optical conveyor belt for single neutral atoms},\ }\href
  {https://doi.org/10.1007/s003400100722} {\bibfield  {journal} {\bibinfo
  {journal} {Applied Physics B}\ }\textbf {\bibinfo {volume} {73}},\ \bibinfo
  {pages} {819} (\bibinfo {year} {2001})}\BibitemShut {NoStop}%
\bibitem [{\citenamefont {Hauck}\ \emph {et~al.}(2021)\citenamefont {Hauck},
  \citenamefont {Alber},\ and\ \citenamefont {Stojanović}}]{hauck_2021}%
  \BibitemOpen
  \bibfield  {author} {\bibinfo {author} {\bibfnamefont {S.~H.}\ \bibnamefont
  {Hauck}}, \bibinfo {author} {\bibfnamefont {G.}~\bibnamefont {Alber}},\ and\
  \bibinfo {author} {\bibfnamefont {V.~M.}\ \bibnamefont {Stojanović}},\
  }\bibfield  {title} {\bibinfo {title} {Single-atom transport in optical
  conveyor belts: {Enhanced} shortcuts-to-adiabaticity approach},\ }\href
  {https://doi.org/10.1103/PhysRevA.104.053110} {\bibfield  {journal} {\bibinfo
   {journal} {Physical Review A}\ }\textbf {\bibinfo {volume} {104}},\ \bibinfo
  {pages} {053110} (\bibinfo {year} {2021})}\BibitemShut {NoStop}%
\bibitem [{\citenamefont {Klostermann}\ \emph {et~al.}(2022)\citenamefont
  {Klostermann}, \citenamefont {Cabrera}, \citenamefont {von Raven},
  \citenamefont {Wienand}, \citenamefont {Schweizer}, \citenamefont {Bloch},\
  and\ \citenamefont {Aidelsburger}}]{klostermann_2022}%
  \BibitemOpen
  \bibfield  {author} {\bibinfo {author} {\bibfnamefont {T.}~\bibnamefont
  {Klostermann}}, \bibinfo {author} {\bibfnamefont {C.~R.}\ \bibnamefont
  {Cabrera}}, \bibinfo {author} {\bibfnamefont {H.}~\bibnamefont {von Raven}},
  \bibinfo {author} {\bibfnamefont {J.~F.}\ \bibnamefont {Wienand}}, \bibinfo
  {author} {\bibfnamefont {C.}~\bibnamefont {Schweizer}}, \bibinfo {author}
  {\bibfnamefont {I.}~\bibnamefont {Bloch}},\ and\ \bibinfo {author}
  {\bibfnamefont {M.}~\bibnamefont {Aidelsburger}},\ }\bibfield  {title}
  {\bibinfo {title} {Fast long-distance transport of cold cesium atoms},\
  }\href {https://doi.org/10.1103/PhysRevA.105.043319} {\bibfield  {journal}
  {\bibinfo  {journal} {Physical Review A}\ }\textbf {\bibinfo {volume}
  {105}},\ \bibinfo {pages} {043319} (\bibinfo {year} {2022})}\BibitemShut
  {NoStop}%
\end{thebibliography}%

\end{document}